\documentclass[journal,twoside,web]{ieeecolor}
\usepackage{generic}
\usepackage{cite}
\usepackage{amsmath,amssymb,amsfonts}

\usepackage{graphicx}
\usepackage{textcomp}
\usepackage[utf8]{inputenc} 
\usepackage[T1]{fontenc}    
\usepackage{hyperref}       
\usepackage{url}            
\usepackage{booktabs}       
\usepackage{amsfonts}       
\usepackage{nicefrac}       
\usepackage{microtype}      
\usepackage{graphicx}
\usepackage{tabularx}
\usepackage{hyperref}
\usepackage{color}
\usepackage{amsmath}

\usepackage{algorithm}  
\usepackage[noend]{algpseudocode} 
\usepackage{subfigure}
\usepackage{multirow}
\usepackage{soul}
\def\BibTeX{{\rm B\kern-.05em{\sc i\kern-.025em b}\kern-.08em
    T\kern-.1667em\lower.7ex\hbox{E}\kern-.125emX}}
\begin{document}
\title{ALNSynergy: a graph convolutional network with multi-representation alignment for drug synergy prediction}
\author{Xinxing~Yang, Jiachen~Li, Xiao~Kang, Guojin~Pei, Keyu~Liu, Genke~Yang, and Jian~Chu \IEEEmembership{}
\thanks{Xinxing Yang is with the Department of Biomedical Engineering, Oregon Health \& Science University, Portland, OR, USA. Email: yangxin@ohsu.edu.}
\thanks{Jiachen Li, Kang Xiao, Guojin Pei, Genke Yang and Jian Chu are with the Department of Automation, Shanghai Jiaotong University, Shanghai 200240, China. Email: kangxiao@sjtu.edu.cn, lijc0804@sjtu.edu.cn; 17601626618@sjtu.edu.cn; gkyang@sjtu.edu.cn; chujian@sjtu.edu.cn.}
\thanks{Keyu Liu is with the School of Computer, Jiangsu University of Science and Technology, Zhenjiang 212100, China. Email: kyliu@just.edu.cn.}
\thanks{Corresponding Author: Xinxing~Yang; Email: yangxin@ohsu.edu.}
}

\maketitle

\begin{abstract}
Drug combination refers to the use of two or more drugs to treat a specific disease at the same time. It is currently the mainstream way to treat complex diseases. Compared with single drugs, drug combinations have better efficacy and can better inhibit toxicity and drug resistance. The computational model based on deep learning concatenates the representation of multiple drugs and the corresponding cell line feature as input, and the output is whether the drug combination can have an inhibitory effect on the cell line. However, this strategy of concatenating multiple representations has the following defects: the alignment of drug representation and cell line representation is ignored, resulting in the synergistic relationship not being reflected positionally in the embedding space. Moreover, the alignment measurement function in deep learning cannot be suitable for drug synergy prediction tasks due to differences in input types. Therefore, in this work, we propose ALNSynergy, a graph convolutional network with multi-representation alignment for predicting drug synergy. In the ALNSynergy model, we designed a multi-representation alignment function suitable for the drug synergy prediction task so that the positional relationship between drug representations and cell line representation is reflected in the embedding space. In addition, the vector modulus of drug representations and cell line representation is considered to improve the accuracy of calculation results and accelerate model convergence. Finally, many relevant experiments were run on multiple drug synergy datasets to verify the effectiveness of the above innovative elements and the excellence of the ALNSynergy model.

\end{abstract}


\section{Introduction}
\label{sec:introduction}
\IEEEPARstart{T}{he} design strategy of "single compound, single target" is followed in the traditional drug research and development process. When faced with complex diseases such as cancer and fungal infections, the therapeutic ability of a single drug will become ineffective during long-term use by the patient. This is because the pathogenesis of these diseases involves multiple targets, and it is difficult for a single drug to comprehensively inhibit the disorder of numerous targets. In addition, patients suffering from the above diseases require long-term use of drugs, which can lead to changes in cellular mechanisms and the development of drug resistance. 

Drug combination refers to using two or more drugs to treat a specific disease simultaneously \cite{drug}. According to research in the relevant literature, when faced with complex diseases, compared with single drugs, drug combination strategies can achieve better efficacy while using fewer drug doses, and toxicity and drug resistance can also be better suppressed. 

Currently, high-throughput screening technology and computational models are mainly used to determine the therapeutic effect of drug combinations on diseases. High-throughput screening techniques use automated instruments to measure the effects of different concentrations of drug combinations on diseased cell lines. However, the huge combination space makes it theoretically unfeasible. The computational model uses modeling technology to learn treatment rules from existing drug synergy samples. It has the advantage of being fast and efficient, which has become the current mainstream prediction method. The high-quality samples accumulated by high-throughput screening technology provide the necessary conditions for the success of computational models.

In general, drug synergy prediction models \cite{a1,a2,a3,a4,a5,a6,a7,a8,a9,10} usually use drug features (molecular fingerprints, molecular graph, etc.) and cell line features (gene expression profiles, etc.) information as model input. The output is the interaction relationship of drug combinations. The synergistic relationship indicates that the combined effect of multiple drugs is greater than the sum of the individual drugs. The effect of antagonistic relationships is opposite to that of synergistic relationships. In the early stages of research, researchers used machine learning models to predict the effects of drug combinations. For example, Liu et al. \cite{r1} used a tree model to predict new synergistic drug combinations. Sałat et al. \cite{r2} used a support vector machine model to predict new drug combinations for use in a mouse model of neuropathic pain.

With the successful application of deep learning in drug repositioning \cite{y1,y2,y3,y8,y4,y9,y5} and drug-target binding affinity prediction \cite{y7,y6}, researchers have begun to use deep learning to model drug synergy prediction problems. Preuer et al. \cite{r3} proposed the DeepSynergy model, which is the first model to apply deep learning to the problem of drug synergy prediction. This model processes the chemical information of two drugs and the genomic information of the disease to form a numerical representation vector and concatenates the three numerical representation vectors into a one-dimensional vector as the model input. The DeepSynergy model inputs this one-dimensional vector into a multi-layer deep neural network to obtain the predicted synergy score. However, the performance of DeepSynergy models is limited by input features and model architecture. Liu et al. \cite{r4} believe that the interaction information between genes and the interaction information between drugs greatly help the prediction of drug synergy. So, they proposed the TranSynergy model. First, the TranSynergy model uses the node information learned on the protein interaction network using the graph embedding algorithm as drug features. Second, gene expression data and gene dependence profiles related to disease mechanisms are employed as cell line features. Finally, the TranSynergy model concatenates drug features and cell line features. It then inputs them into the dimensionality reduction module, self-attention transformer, and fully connected network in sequence to obtain the predicted synergy score. The above drug representations of DeepSynergy and TranSynergy models do not utilize the structural information of molecular graphs, while graph neural networks have strong structural learning capabilities on graph data. Therefore, Wang et al. \cite{r5} proposed the DeepDDS model and applied the graph neural network \cite{gcn1,gcn2} to the problem of drug synergy prediction. The DeepDDS model uses a graph convolutional network to learn drug representations from molecular graphs and a multi-layer neural network to learn cell line representations from genomic information. Then, the DeepDDS model concatenates the drug representation and the cell line representation into a one-dimensional vector, which is input into the fully connected network to obtain the predicted synergy score.

When two drugs are combined to treat a disease, previous work concatenated the representation of the first drug, the representation of the second drug, and the representation of the cell line into a fully connected neural network to obtain the prediction value. In this computing structure, the dimensions of drug representation and cell line representation can be inconsistent, so the drug and cell line representations do not appear in the same embedding space. This approach fails to take into account the similarities and correlations between drug representation and cell line representation. What is particularly important is that it ignores the alignment between representations, and the positional relationship between drug representations and cell line representations that have a synergistic relationship (antagonism relationship) cannot be reflected in the embedding space.

Representation alignment originates from contrastive learning, which mainly measures the similarity of entities in different domains in the embedding space. By optimizing the representation alignment, samples from two domains can be mapped into the same embedding space so that the positions of similar samples in this space are approximated. For example, in recommendation systems, users and items belong to their respective spatial domains. Alignment refers to the similarity of representations between interactive users and items. Users and items with interactive relationships can have approximate representations in the same embedding space by optimizing representation alignment. Therefore, in essence, representation alignment can reflect the quality of representation, which is of great help to improve the prediction accuracy of the model.

The input of the previous representation alignment function is a binary group (user, product), etc., which cannot be applied to the drug synergy prediction problem where the input is a triplet (drug one, drug two, cell line). Representation alignment in the traditional sense is to calculate the mean square error between two representation vectors. The smaller the difference, the higher the alignment between the two. The alignment of triples requires pairing operations between representations, and the mean square error of various combinations is calculated in turn. This calculation method is cumbersome and does not consider the relationship between drug combinations and corresponding cell lines. Therefore, how to introduce the concept of alignment into the problem of drug synergy prediction is a challenging task.

In this work, we propose ALNSynergy, a graph convolutional network with multi-representation alignment for predicting drug synergy. In the ALNSynergy, we draw on representation alignment in contrastive learning \cite{t1,t2,t3} and design an alignment function suitable for drug synergy prediction, which measures the representation alignment between triples (drug one, drug two, cell line). The alignment function first uses the Hadamard product to perform an element-wise multiplication operation on the three representation vectors. Then, it performs a linear addition of each element in the product result to obtain the alignment between the three representations. Specifically, the representation vectors of the first drug and the second drug are first subjected to a Hadamard product operation so that the elements of each dimension are multiplied to form a drug combination representation vector. This corresponds to a certain extent with the concept of drug synergy. Then, the drug combination representation vector and the cell line representation vector are subjected to an inner product operation to obtain the similarity (alignment) between the two. In addition, we directly optimized the representation alignment based on the original optimization function so that the alignment between the triplets with synergistic relationships was as large as possible. The alignment of the triplets with antagonistic relationships is as small as possible, so the positional relationship between the drug representation vector and the cell line representation vector with a synergistic relationship (antagonism relationship) is reflected in the embedding space.

\begin{figure*}[t]
	\centering
	\includegraphics[scale=0.6]{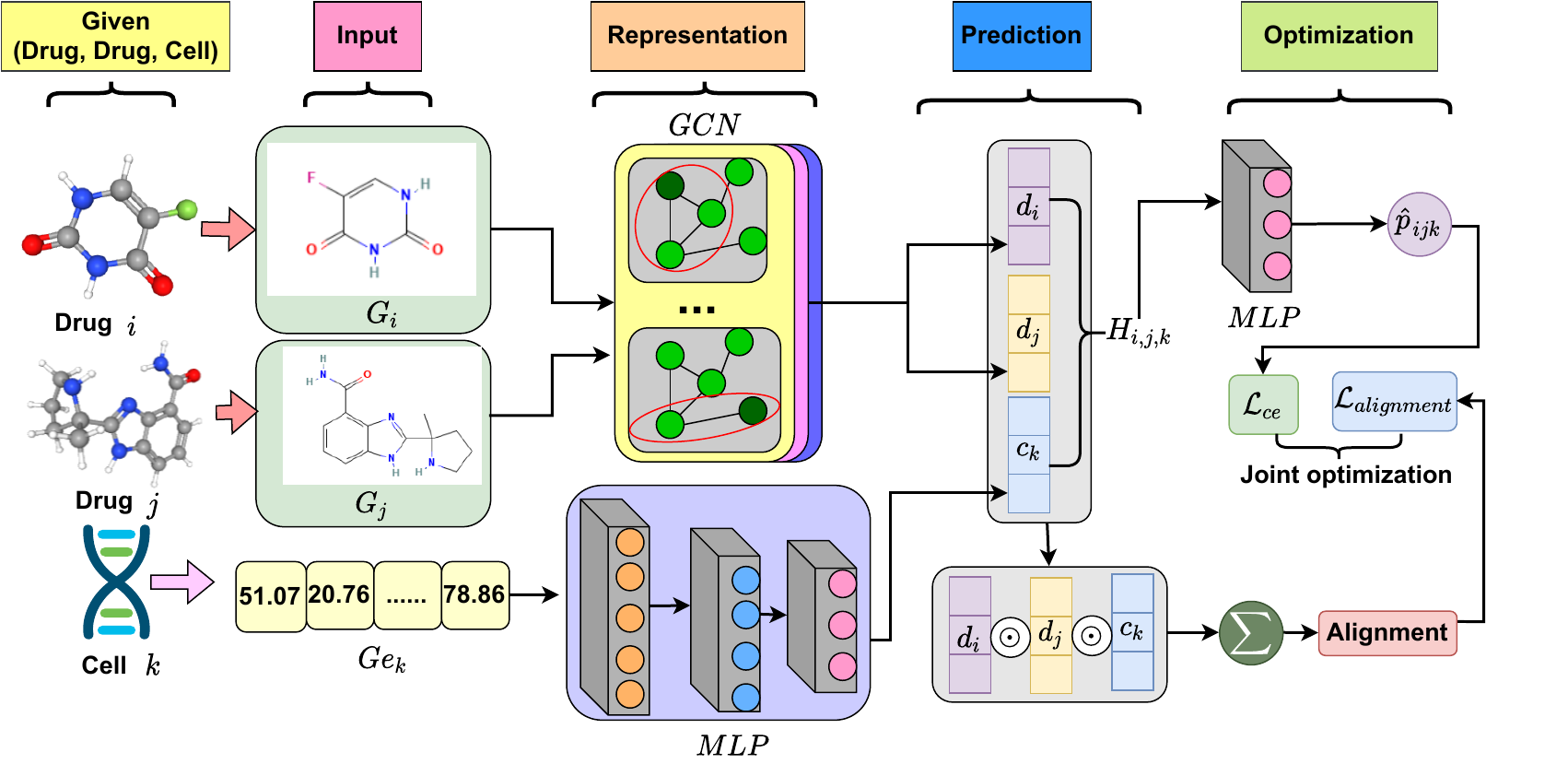}
	\caption{The architecture of the ALNSynergy model}
	\label{}
\end{figure*}

However, as a similarity measure, the inner product does not explicitly consider the module of the three representation vectors, which will affect the accuracy of similarity calculation and the convergence of the model. Regarding the accuracy of similarity calculation, Since more popular drugs or cell lines contribute significantly to the model's training, its module length becomes very large. The nature of the inner product will be affected by the module length. Suppose the module lengths of the three representation vectors (drug one, drug two, and cell line) differ significantly. In that case, it will lead to large similarity values among the three unrelated ones, thus affecting the accuracy of the calculation.
Regarding convergence: The representation vector module lengths of various magnitudes damage the convergence of the model. The model still cannot show a performance convergence trend after multiple iterations. This wastes a lot of time and computing resources. Therefore, based on the above alignment function, we introduce a temperature coefficient to control the variation range of the module length to ensure that multiple module lengths are of the same magnitude. Specifically, we normalize the module lengths of the three representation vectors. The temperature coefficient is then used to scale the module length to observe the impact of different magnitudes on the model. The advantage of this approach is that the module length is considered in the loss function so that the length is compressed to the same magnitude and can increase the accuracy of the calculation.

The three contributions of this work are summarized below.

\begin{enumerate}
	\item We designed a representation alignment function suitable for drug synergy prediction to reflect the positional relationship between the drug representations and the cell line representation in the embedding space.
	\item We additionally consider the representation module in the above alignment function to improve the computational similarity and speed up the convergence of the model.
	\item Relevant experiments were run on multiple drug synergy data sets to verify the effectiveness of the above innovative elements and the excellence of the ALNSynergy model.
\end{enumerate}

The architecture of the ALNSynergy model and computational details are introduced in Section 2. The dataset, experimental setup, ablation experiments of innovative elements, and comparison experiments with state-of-the-art models are presented in Section 3. Section 4 summarizes the work of this paper and the directions for future improvements.

\section{Method}

Figure 1 shows the architecture of the ALNSynergy model. The architecture consists of three components: the drug/cell line input layer, the representation learning layer, and the prediction layer. Given a triplet $(i,j,c)$, the ALNSynergy model first uses the molecular graph as the input of drug $i$ and drug $j$, and the genomic data as the input of cell line $c$. Subsequently, the ALNSynergy model inputs molecular graphs and genomics data into graph convolutional networks and multi-layer perceptrons to learn their respective representations. Finally, the ALNSynergy model concatenates the three representations and inputs them into a multi-layer, fully connected network's prediction layer to obtain the synergy score. In the prediction layer, we designed an alignment function to calculate the alignment between the representation of drug $i$, the representation of drug $j$, and the representation of cell line $c$. To play the role of multi-representation alignment, the prediction loss and alignment loss are optimized simultaneously in the parameter optimization stage. The above modules will be introduced in detail below.

\subsection{Drug representation learning module}

Given $(i,j,c)$, the ALNSynergy model first uses a graph convolutional network to perform representation learning on drugs $i$ and $j$. Taking drug $i$ as an example, its molecular graph is $G=(X,A)$, $X\in \mathbb{R}^{n \times k}$ represents the atom set, where each row represents the $k$ features of the corresponding atom. $A\in \mathbb{R}^{n \times n}$ represents a set of bonds, where each value represents whether there is a bond between two atoms. Subsequently, the ALNSynergy model inputs $G$ into a graph convolutional network for learning the representation of drug $i$. The propagation rule of the graph convolutional network is shown in equation (1).

\begin{equation}
	Z^{l+1} = \sigma(\tilde{D}^{-\frac{1}{2}}\tilde{A} \tilde{D}^{-\frac{1}{2}} Z^{l}W^{l})
\end{equation}

Among them, $\tilde{A}=A+I$ means adding a self-connected adjacency matrix. $\tilde{D}_{ii}=\sum_{j}\tilde{A}_{ij}$. $Z^{l+1}$ represents the output of layer $l$, $Z^{l}$ represents the input of layer $l$, and $W^l$ represents the parameters of layer $l$. $Z^{0}$ is $X$. $\sigma$ is the activation function.

Then, after the calculation of $l$ graph convolutional layers, the ALNSynergy model uses equation (2) to convert the node-level representation $Z_{l+1}$ into the graph-level representation $d_i$ with $h$ dimensions. 

\begin{equation}
	d_i = GMP(Z^{l+1})
\end{equation}

$GMP$ represents the global max pooling operation. For drug $j$, using the same operation as above to obtain its representation $d_j$ with $h$ dimensions.

\subsection{Cell line representation learning module}

For the cell line $k$, the ALNSynergy model uses the genome information $Ge \in \mathbb{R}^{e \times 1}$ as its input, and $e$ represents the dimension of the genome information. To learn its representation, the ALNSynergy model feeds it into a multilayer perceptron model, as shown in equation (3).

\begin{equation}
	c_k = W_3\sigma(W_2 \sigma(W_1G+b_1) + b_2)+b_3
\end{equation}

The multilayer perceptron model consists of three hidden layers. $W_1 \in \mathbb{R}^{h^1 \times e}$, $W_2 \in \mathbb{R}^{h^2 \times h^1}$ and $W_3 \in \mathbb{R}^{h \times h^2}$ are the weight parameters. $b_1\in \mathbb{R}^{h^1 \times 1}$, $b_2\in \mathbb{R}^{h^2 \times 1}$ and $b_3\in \mathbb{R}^{h \times 1}$ are the bias parameters of the corresponding hidden layer respectively. $c_k\in \mathbb{R}^{h \times 1}$ represents the representation vector of cell line $k$.

\subsection{Prediction module}

Then, as shown in equation (4), the ALNSynergy model concatenates $d_i$, $d_j$ and $c_k$ to form a one-dimensional head node vector $H_{i,j,k} \in \mathbb{R}^{3h \times 1}$ according to the relevant rules, and use it as the input of the prediction layer.

\begin{equation}
	H_{i,j,k}=concentrate[d_i,d_j,c_k]
\end{equation}

The ALNSynergy model inputs $H_{i,j,k}$ into the prediction function shown in equation (5) and obtains the synergy score $hat{p}_{ijc} $ of the combination of drug $i$ and drug $j$ on cell line $k$.

\begin{equation}
	\hat{p}_{ijc} = \sigma(V H_{i,j,k}+b)
\end{equation}

As a binary classification problem, this section uses the cross-entropy loss function to optimize the parameters in the ALNSynergy model. The loss function is shown in equation (6).

\begin{equation}
	\mathcal{L}_{ce} = -\big[ {p}_{ijc}  \log \hat{p}_{ijc}  + (1-{p}_{ijc}) \log (1-\hat{p}_{ijc}) \big] 
\end{equation}

Then, the Adam optimization algorithm is used to minimize $L_{bce}$, and the parameters in the ALNSynergy model can be optimized.

\begin{figure}[h]
	\centering
	\includegraphics[scale=0.7]{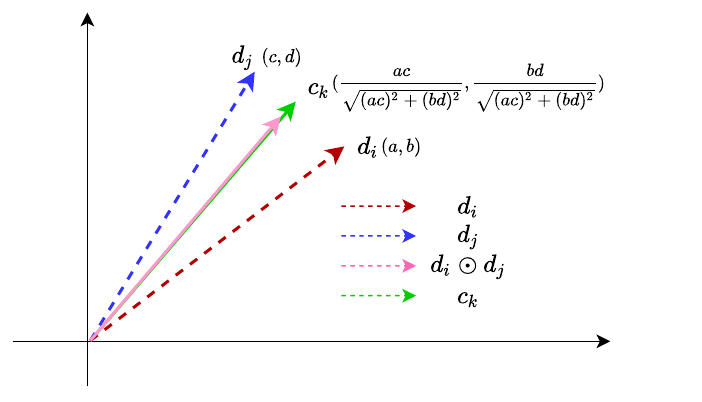}
	\caption{Example of alignment function in two-dimensional space when temperature coefficient is 1}
	\label{}
\end{figure}

\subsection{Multi-representation alignment function}

In order to improve the representation quality of drug $i$, drug $j$ and cell line $k$, so that the synergistic relationship between the three can be reflected positionally in the embedding space, the ALNSynergy model designed an alignment function to measure the alignment between $d_i$, $d_j$ and $c_k$. As shown in equation (7), the ALNSynergy model first runs the Hadamard product operation on $d_i$, $d_j$, and $c_k$ and then linearly adds each element to obtain the alignment between the three representation vectors.

\begin{align}
Alignment_{ijc} &= \sum_{t=1}^{h} d_{it} \odot d_{jt} \odot c_{kt} \\ \notag
&=  (d_i \odot d_j)^T c_k 
\end{align}

Where $\odot $ represents the Hadamard product, which multiplies the elements of the same dimension in the vector. $Alignment_{ijc}$ represents the degree of alignment between the three representation vectors. The multi-representation alignment function can be interpreted from the following perspectives. The representation vectors of drug $i$ and drug $j$ are first subjected to the Hadamard product operation, and the elements of each dimension are multiplied to form the drug combination representation vector $d_i \odot d_j$. This corresponds to the concept of drug combination to a certain extent. Subsequently, the drug combination representation vector $d_i \odot d_j$ and the cell line representation vector $c_k$ are subjected to the inner product operation to obtain the similarity (alignment) between the two representation vectors. If the combination of drug $i$ and drug $j$ can produce a synergistic effect on cell line $k$, the multi-representation alignment function will calculate a larger alignment value, otherwise a smaller alignment value will be calculated.

Subsequently, the ALNSynergy model optimizes the representation alignment based on the original loss function in equation (6) to make the multi-representation alignment function work. The optimization operation is shown in equation (8).

\begin{align}
\mathcal{L} &=  \mathcal{L}_{ce} + \alpha \mathcal{L}_{alignment} \\ \notag
\mathcal{L}_{alignment} &= \frac{1}{N} \sum_{(i,j,k)\in \mathcal{T}} (p_{ijk}-\psi(Alignment_{ijc}))^2
\end{align}

$\psi$ represents the Sigmoid function, which maps $Alignment_{ijc}$ to the $(0,1)$ interval. $\mathcal{L}_{alignment}$ represents the alignment loss. By optimizing this loss, the alignment between $d_i$, $d_j$ and $c_k$ with synergistic relationships can be as large as possible, while the alignment between $d_i$, $d_j$ and $c_k$ with antagonistic relationships can be as small as possible. The drug synergistic relationship is reflected in the embedding space, thereby improving the quality and robustness of drug and cell line representation, and the generalization ability of the ALNSynergy model is enhanced. $\alpha$ is used to adjust the weight of the alignment loss in the final loss function $\mathcal{L}$.

The reason why this alignment function works is that it restricts the dimensions of $d_i$, $d_j$ and $c_k$ to be consistent, and the three vectors exist in the same embedding space. The same embedding space can take into account the similarity and correlation between the drug representation vector and the cell line representation vector. In addition, the alignment function can measure the similarity of entities in two different domains, drugs and cell lines, in the same embedding space. By optimizing the representation alignment, $d_i$, $d_j$, and $c_k$ with synergistic relationships can have approximate representations in the same embedding space, thereby improving the quality of drug representation and cell line representation. Therefore, representation alignment can essentially reflect the quality of representation. Optimizing alignment can enhance the quality and robustness of drug representation and cell line representation, which is of great help in improving the prediction accuracy of the model. In other words, the alignment loss can be used as a regular term in the overall loss function to control the complexity of the representation vector and avoid overfitting of the model.

\subsection{Multi-representation alignment function with adaptive module length}

The above-mentioned multi-representation alignment loss can improve the model's performance. Still, the inner product, as a similarity measure, does not explicitly consider the module length of the three representation vectors, which will affect the accuracy of the similarity calculation results and the convergence of the model. Therefore, this section further optimizes the above-mentioned multi-representation alignment function. As shown in equation (9), based on the above-mentioned multi-representation alignment function, this section introduces a temperature coefficient to control the variation range of the representation vector modulus length to ensure that the modulus lengths are of the same magnitude.

\begin{equation}
Alignment_{ijc}  = \frac{(d_i \odot d_j)^T c_k }{ ||d_i \odot d_j|| ||c_k|| } . \frac{1}{\tau} 
\end{equation}

Specifically, the above alignment function first normalizes the module lengths of the drug combination representation vector and the cell line representation vector. The representation is then scaled using the temperature coefficient $\tau$ to observe the impact of different magnitudes on model performance. The advantage of this approach is that the module length of the representation vector is taken into account in the loss function so that the module length is compressed to the same magnitude, the convergence speed of the model is accelerated, and the accuracy of the similarity value is increased.

Figure 2 shows an example of the alignment function in two dimensions when the temperature coefficient is 1. This alignment function calculates the cosine value of the drug combination representation vector and the cell line representation vector. Since the drug combination representation vector and the cell line representation vector are both normalized vectors, the cosine value is maximum when they are equal (the angle is 0). Subsequent experiments show that this alignment function has better convergence speed.

\section{Experiments and Discussion}

\begin{table}[h]
	\caption{Statistical results of benchmark dataset and independent dataset}
	\centering
	\setlength{\tabcolsep}{3mm}
	\begin{tabular}{@{}ccc@{}}
		\toprule
		Datasets/Statistical results& Benchmark dataset & Independent dataset  \\ \midrule
		Drugs       & 36   & 57    \\
		Cells         & 31    & 24   \\
		Samples  & 12415 & 668 \\ \bottomrule
	\end{tabular}
\end{table}

\subsection{Datasets}

The datasets used in this section are similar to the work of Wang et al. \cite{r5}, a benchmark dataset and an independent dataset. Table 1 lists the statistical results of the two datasets. The sample format of the benchmark dataset is [drug one, drug two, cell line, label value], where SMILES strings represent drug one and drug two, and a naming code represents the cell line. When the label value is 1, the combination of drug one and drug two synergistically affects the cell line. When the label value is 0, the combination of drug one and drug two has an antagonistic effect on the cell line. This benchmark dataset labels samples with a Loewe Additivity Score greater than ten as synergistic samples and samples with a score less than 0 as antagonistic samples. The final benchmark dataset comprised 12,415 samples, including 36 anticancer drugs and 31 human cancer cell lines. In addition, an independent dataset was provided by AstraZeneca, which contains 57 drugs, 24 cell lines, and 668 samples.

\subsection{Experimental settings and evaluation metrics}

We use a five-fold cross-validation strategy for the benchmark dataset to evaluate the model's performance. The 12,415 samples were divided into five parts, four of which were used as the training set, and the remaining one was used as the test set. Finally, the five performances are accumulated and averaged to obtain the final performance of the model. For independent datasets, we train the model on the benchmark dataset and then test the trained model on the independent dataset.

Drug synergy prediction is modeled as a binary classification problem in this section, so the evaluation indicators use AUC (Area Under Curve Area), AUPR (Area Under Precision-Recall Curve), and ACC (Accuracy).

\begin{table}[h]
	\caption{Hyperparameter settings}
	\centering
	\setlength{\tabcolsep}{0.0001mm}
	\begin{tabular}{@{}ccc@{}}
		\toprule
		Hyperparameter                  & Variation range               & Default \\ \midrule
		Drug/cell line representation dimensions                    & $[32,64,128,256,512]$ & 128        \\
		Loss weight $\alpha$                       & $[0,0.01,0.1,0.5,1]$          & 0.5           \\
		Temperature coefficient $\delta$                       & $[0.01,0.1,0.5,1]$               & 0.1             \\
		The number of GCN layers    	& $[1,2,3,4,5]$           & 3           \\
		The number of FCN layers  & $[1,2,3]$            & 3           \\
		Batch size & $[64,128,256]$      & 128           \\
		Learning rate   & $[0.0001,0.0005,0.001,0.005]$      & 0.0001           \\ \bottomrule
	\end{tabular}
\end{table}

\subsection{Hyperparameter settings}

The hyperparameter settings in the ALNSynergy model are shown in Table 2, including drug/cell line representation dimensions, loss weight $\alpha$, temperature coefficient $\delta$, the number of graph convolution layers, and the number of fully connected network layers, batch size, and learning rate. The variation range of drug/cell line representation dimensions is $[32,64,128,256,512]$. The variation range of $\alpha$ is $[0,0.01,0.1,0.5,1]$. The variation range of $\delta$ is $[0.01,0.1,0.5,1]$. The variation range of graph convolutional layers is $[1,2,3,4,5]$. The fully connected network layer variation range is $[1,2,3]$. The variation range of batch size is $[64,128,256]$. In the Adam optimizer, the change interval of the learning rate is $[0.0001,0.0005,0.001,0.005]$. The values of the above hyperparameters are determined by their performance on the validation set, and the default values are 128, 0.5, 0.1, 3, 3, 128, and 0.0001, respectively.

\subsection{Effectiveness of multi-representation alignment function}

This section runs an ablation experiment of the multi-representation alignment function shown in equation (7) to verify the effectiveness of this innovative element. In equation (7), a multi-representation alignment function suitable for drug synergy prediction is used to improve the quality of drug representation and cell line representation, and the alignment function is made effective through the optimization strategy in equation (8). Parameter $\alpha$ is used to adjust the weight of multi-representation alignment loss in the model optimization process. Specifically, when $\alpha$ is 0, the multi-representation alignment function does not participate in the optimization process of model parameters. Only when $\alpha$ is greater than 0 the loss function will take multi-representation alignment into account. Therefore, the following ablation experiments are conducted by adjusting the value of $\alpha$ in the interval $[0,1,2]$ to observe the rising and falling trends in the performance of the ALNSynergy model.

\begin{table}[h]
	\caption{Experimental results of ALNSynergy model under different $\alpha$ values}
	\centering
	\setlength{\tabcolsep}{4mm}
	\begin{tabular}{@{}ccccc@{}}
		\toprule
		Dataset & $\alpha $     & AUC   & AUPR    & ACC \\ \midrule
		Benchmark dataset   & value=0    & 0.92 & 0.92 & 0.84   \\
		& value=1    & 0.93 & 0.92 & 0.85   \\ 
		& value=2    & \textbf{0.93} & \textbf{0.93} & \textbf{0.85 }  \\ \bottomrule
	\end{tabular}
\end{table}

\begin{figure}[h]
	\centering
	\subfigure[Trend]{
		\includegraphics[width=4cm,height=3.2cm]{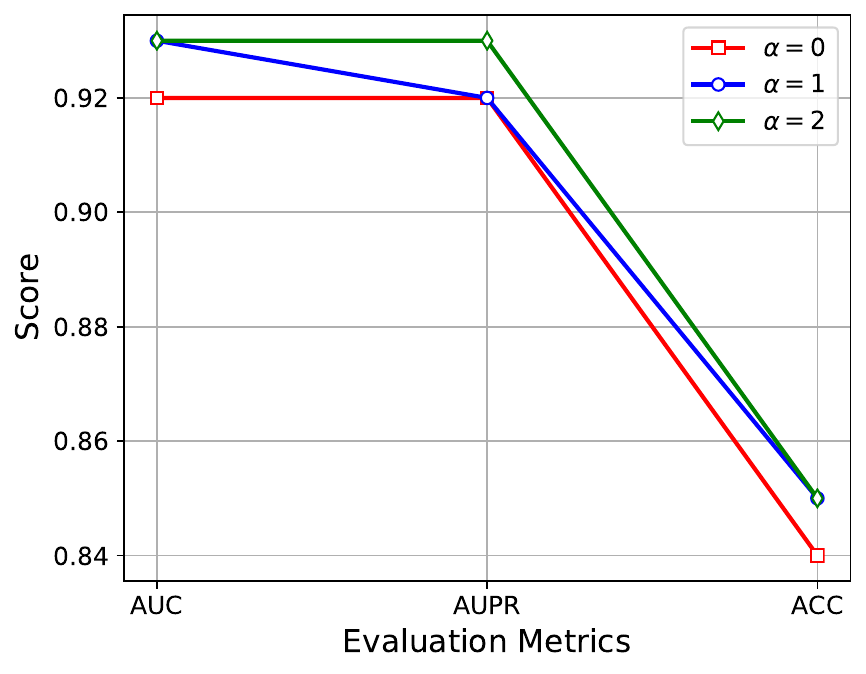}
	}
	\subfigure[Improvement]{
		\includegraphics[width=4cm,height=3.2cm]{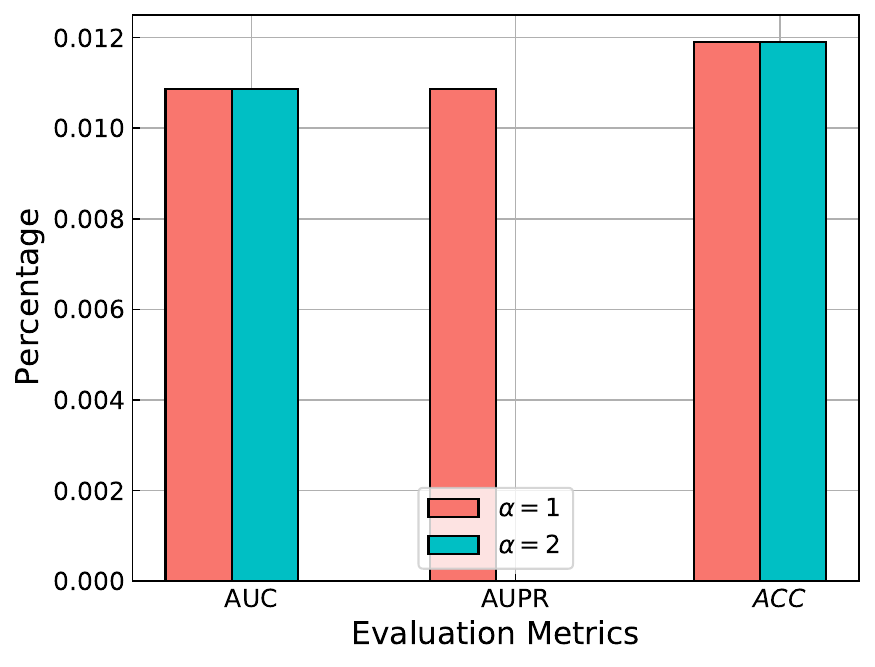}
	}
	
	\caption{The changing trend and improvement percentage of the ALNSynergy model under different $\alpha$ values}
\end{figure}

The experimental results of the ALNSynergy model under different $\alpha$ values are shown in Table 3. The changing trend of experimental performance is shown in Figure 3(a). Compared with the ALNSynergy model at $\alpha=0$, its improvement percentage at $\alpha=1$ and $\alpha=2$ is shown in Figure 3(b). As shown in the Figure, optimizing the representation quality through the multi-representation alignment function can improve the ALNSynergy model's performance. Specifically, when $\alpha$ is 0, that is, when the multi-representation alignment function does not optimize the representation quality, the ALNSynergy model achieves the worst prediction performance, and its scores under the AUC, AUPR, and ACC evaluation metrics are respectively are 0.92, 0.92 and 0.84.

When the values of $\alpha$ are 1 and 2, the prediction performance of the ALNSynergy model is better. Especially when the value of $\alpha$ is 2, the scores of the ALNSynergy model under the AUC, AUPR, and ACC evaluation metrics are 0.93, 0.93, and 0.85, respectively. Compared with $\alpha$ being 0 when $\alpha$ is 2, the ALNSynergy model has improved by 1.1\%, 1.1\%, and 1.2\%, respectively, under the three evaluation metrics.

In addition, it should be noted that the overall performance of the ALNSynergy model has been improving as the $\alpha$ value increases. When the $\alpha$ value is 1, the AUPR evaluation metrics value of the ALNSynergy model is the same as the ALNSynergy model when the $\alpha$ value is 0, while the AUC and ACC evaluation metrics have increased by 1.1\% and 1.2\% respectively.

From the analysis of the above experimental results, it can be concluded that the multi-representation alignment function in equation (7) can be used to improve the quality of drug representation and cell line representation, thereby improving the performance of the ALNSynergy model. This validates the effectiveness of this innovative element. However, it should be noted that the value of the loss weight $\alpha$ needs to be set carefully.

\subsection{Effectiveness of multi-representation alignment function with adaptive module length}

This section runs the ablation experiment of the multi-representation alignment function with adaptive module length in equation (9) to verify whether this alignment function is better than the alignment function that does not consider the module length. The difference between the multi-representation alignment function in equation (9) and equation (7) is that the former introduces a temperature coefficient to control the variation range of the module length to ensure that the module length is of the same magnitude, thus increasing the accuracy of similar values and the convergence of the model. Therefore, this section compares the experimental results of ALNSynergy-WM, ALNSynergy-NM, and ALNSynergy-NA models. Among them, the ALNSynergy-WM model uses the multi-representation alignment function shown in equation (9), considering the representation module length. The ALNSynergy-NM model uses the multi-representation alignment function shown in equation (7). That is, the representation module length is not considered. The ALNSynergy-NA model does not use the multi-representation alignment function. That is, the value of parameter $\alpha$ is zero. The purpose of comparing the above three models is two-fold: the first is to verify whether the multi-representation alignment function with adaptive module length can improve the model prediction performance; the second is to verify whether it can enhance the convergence of the model.

\begin{table}[h]
	\caption{Experimental results of ALNSynergy-NA, ALNSynergy-NM and ALNSynergy-WM mdoels}
	\centering
	\setlength{\tabcolsep}{2mm}
	\begin{tabular}{@{}ccccc@{}}
		\toprule
		Dataset & Model     & AUC   & AUPR    & ACC \\ \midrule
		Benchmark dataset   & ALNSynergy-NA    & 0.92 & 0.92 & 0.84   \\
		& ALNSynergy-NM    & 0.93 & 0.93 & 0.85   \\ 
		& ALNSynergy-WM    & \textbf{0.93} & \textbf{0.93} & \textbf{0.86}  \\ \bottomrule
	\end{tabular}
\end{table}

\begin{figure}[h]
	\centering
	\subfigure[Trend]{
		\includegraphics[width=4cm,height=3.2cm]{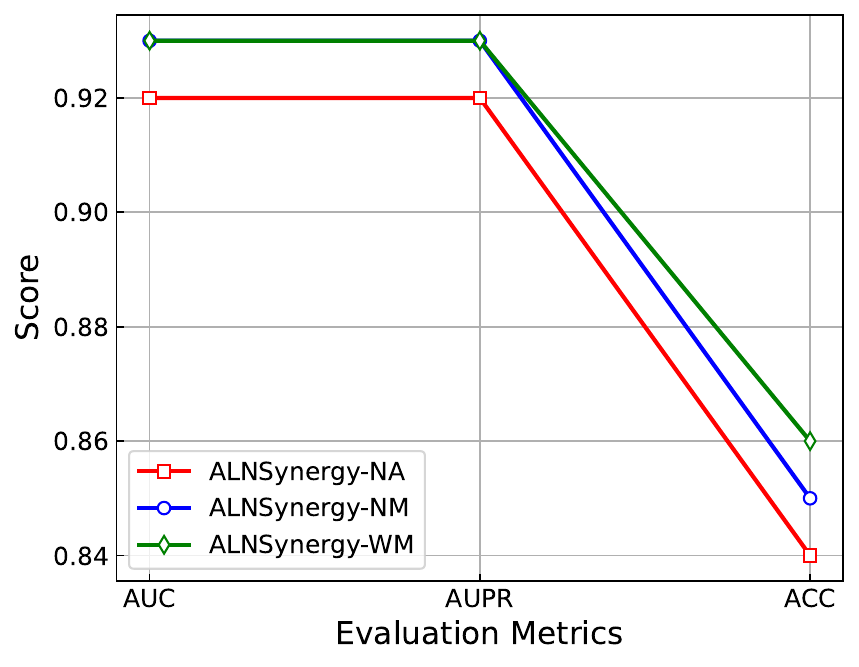}
		
	}
	\subfigure[Improvement]{
		\includegraphics[width=4cm,height=3.2cm]{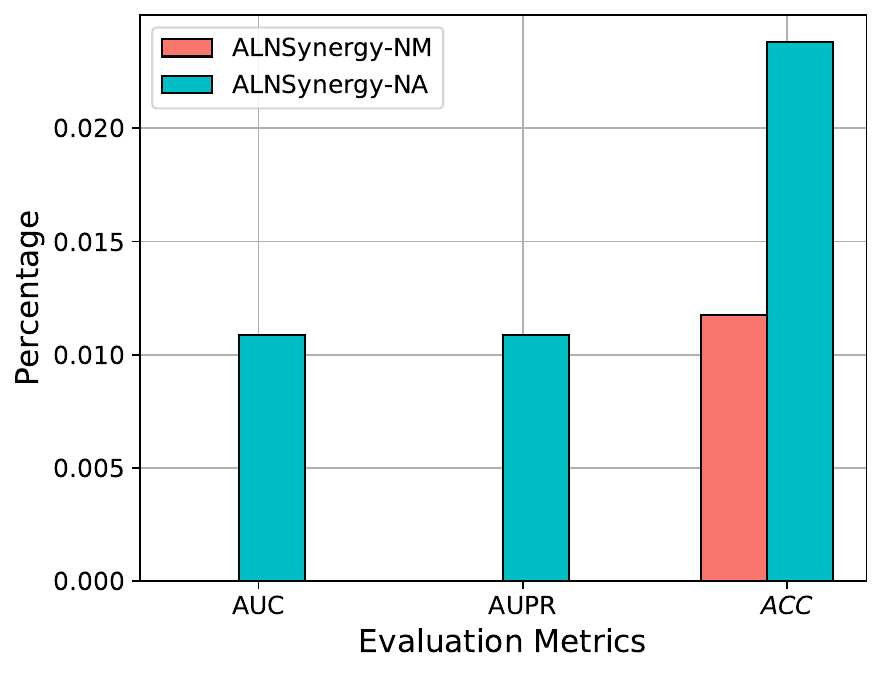}
	}
	
	\caption{The changing trend and improvement percentage of the ALNSynergy-NA model, ALNSynergy-NM model and ALNSynergy-WM model}
\end{figure}

The experimental results of the ALNSynergy-WM model, ALNSynergy-NM model, and ALNSynergy-NA model are shown in Table 4. The experimental performance change trend among the three models is shown in Figure 4 (a). Compared with the ALNSynergy-NA model and the ALNSynergy-NM model, the improvement percentage of the ALNSynergy-WM model under the three evaluation metrics is shown in Figure 4 (b).
First, the ALNSynergy-WM model achieved the optimal prediction performance, with AUC, AUPR, and ACC values of 0.93, 0.93, and 0.86 respectively. Compared with the ALNSynergy-NA model, the improvement values of the ALNSynergy-WM model in the three evaluation metrics are 1.1\%, 1.1\%, and 2.4\%, respectively. Therefore, the multi-representation alignment function with adaptive module length helps improve the model's prediction performance. Compared with the ALNSynergy-NM model, the ALNSynergy-WM model only improved the ACC evaluation metric by 1.2\%, while the AUC and ACUPR metrics remained unchanged. The experimental results show that compared with the multi-representation alignment function shown in equation(7), the multi-representation alignment function with adaptive module length shown in equation (9) can further improve the model's prediction performance, but the improvement effect is limited.

\begin{figure*}[h]
	\centering
	\subfigure[AUC]{
		\includegraphics[width=5cm,height=4cm]{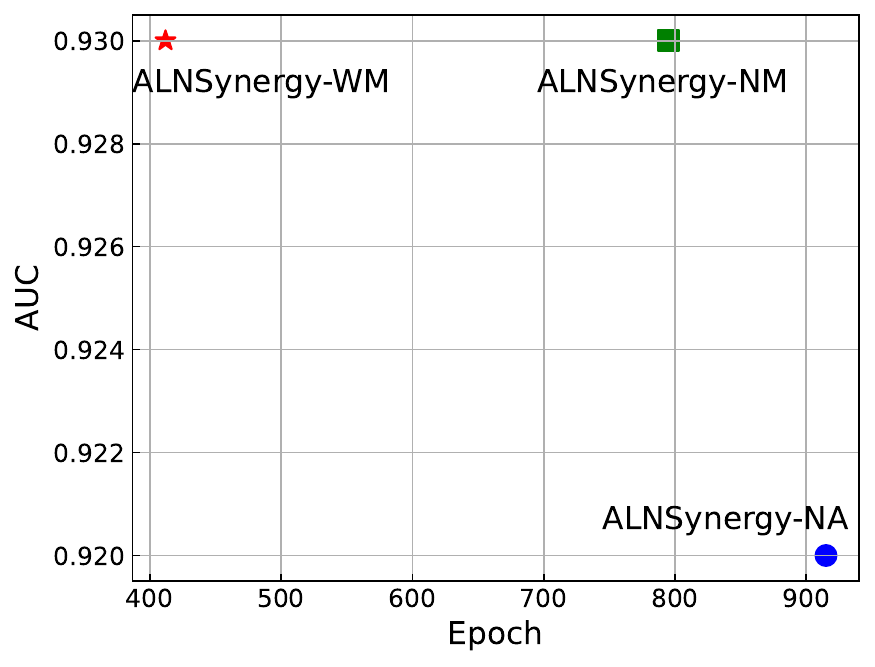}
		
	}
	\quad
	\subfigure[AUPR]{
		\includegraphics[width=5cm,height=4cm]{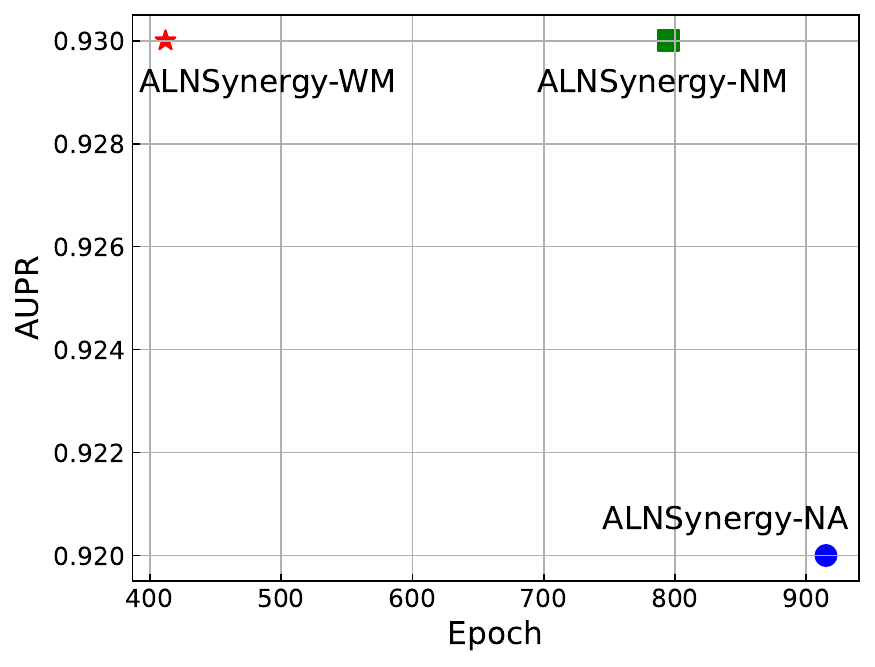}
	}
	\quad
	\subfigure[ACC]{
		\includegraphics[width=5cm,height=4cm]{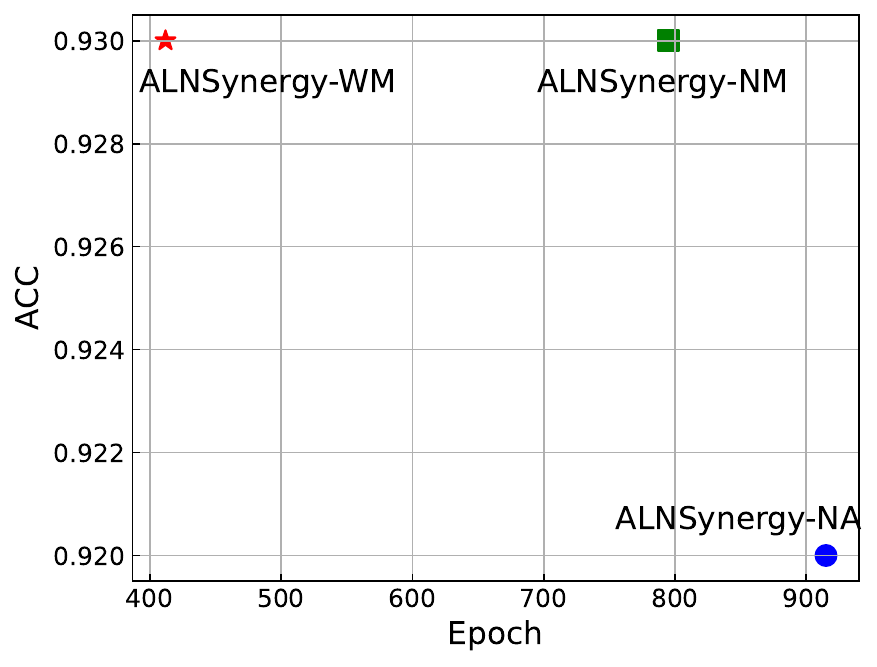}
	}
	
	\caption{The convergence speed of ALNSynergy-WM model, ALNSynergy-NM model and ALNSynergy-NA model}
\end{figure*}

Next, this section analyzes the impact of the multi-representation alignment function with adaptive module length on model convergence. This section treats the number of iterations of the model as time spent. The smaller the number of iterations, the faster the convergence speed. Figure 5 presents the convergence speed of the ALNSynergy-WM model, ALNSynergy-NM model, and ALNSynergy-NA model. First, the ALNSynergy-WM model can reach the convergence state after 412 iterations. The ALNSynergy-NM model and ALNSynergy-NA model require 795 and 915 iterations to reach the convergence state. Compared with the ALNSynergy-NM model and ALNSynergy-NA model, the convergence speed of the ALNSynergy-WM model has increased by 55.0\% and 48.2\%, respectively, greatly speeding up the model training process.

This section successfully verifies that the multi-representation alignment function with adaptive module length shown in equation (9) has better performance and convergence through the analysis and discussion of the above experimental results. This is mainly because the temperature coefficient adjusts the representation mode length, giving it better calculation accuracy and faster convergence speed. Therefore, the above conclusion verifies the effectiveness of the adaptive module length shown in equation (9).

\subsection{Comparison experiment}

\begin{itemize}
	\item \textbf{Support Vector Machine \cite{svm} (SVM):} The SVM model distinguishes drug combinations with synergistic or antagonistic relationships by maximizing the interval.
	\item \textbf{Multilayer Perceptron \cite{mlp} (MLP):} The MLP model concatenates drug combination features and cell line features into a multilayer neural network to obtain a predicted synergy score.
	\item \textbf{Random Forest \cite{rf} (RF):} The RF model uses multiple decision trees to predict drug synergy problems.
	\item \textbf{DeepSynergy \cite{r3}:} The DeepSynergy model processes the chemical information of the drug and the genomic data related to the disease to form a numerical vector and concatenates the three representation vectors into a one-dimensional vector. The DeepSynergy model then inputs this one-dimensional vector into a multilayer deep neural network to obtain the predicted synergy score.
	\item \textbf{DeepDDS \cite{r5}:} The DeepDDS model uses a graph convolutional network to learn drug representations from molecular graphs and a multilayer neural network to learn cell line representations from genomic information. Next, the model concatenates the drug and cell line representations into a one-dimensional vector, which is then input into a fully connected network to obtain the predicted synergy score.
\end{itemize}

\begin{table}[h]
	\caption{Experimental results of ALNSynergy and comparison mdoels on benchmark dataset}
	\centering
	\setlength{\tabcolsep}{7mm}
	\begin{tabular}{@{}cccc@{}}
		\toprule
		\textbf{Method/Metric} & \textbf{AUC} & \textbf{AUPR} & \textbf{ACC} \\ \midrule
		ALNSynergy                 & \textbf{0.93}         &\textbf{ 0.93}          & \textbf{0.86}              \\
		DeepDDS                & 0.92         & 0.92          & 0.84              \\
		DeepSynergy            & 0.88         & 0.87          & 0.80              \\
		RF			           & 0.86         & 0.85          & 0.77              \\
		MLP                    & 0.65         & 0.63          & 0.56              \\
		SVM                    & 0.58         & 0.56          & 0.54              \\ \bottomrule
	\end{tabular}
\end{table}

\begin{table}[h]
	\caption{Experimental results of ALNSynergy and comparison mdoels on independent dataset}
	\centering
	\setlength{\tabcolsep}{7mm}
	\begin{tabular}{@{}cccc@{}}
		\toprule
		\textbf{Method/Metric} & \textbf{AUC} & \textbf{AUPR} & \textbf{ACC} \\ \midrule
		ALNSynergy                 & \textbf{0.66}         & \textbf{0.83}          & \textbf{0.63  }            \\
		DeepDDS                & 0.65         & 0.82          & 0.62              \\
		DeepSynergy            & 0.55         & 0.71          & 0.47              \\
		RF          		   & 0.53         & 0.76          & 0.50              \\
		MLP                    & 0.53         & 0.74          & 0.53              \\
		SVM                    & 0.47         & 0.71          & 0.54              \\ \bottomrule
	\end{tabular}
\end{table}

Table 5 shows the experimental results of the ALNSynergy model and the above comparison models on the benchmark dataset. As shown in the Table 5, the ALNSynergy model outperforms previous deep learning models and machine learning models. Specifically, the three evaluation metrics values of the ALNSynergy model are 0.93, 0.93, and 0.86, respectively. The machine learning model with the best performance is RF, whose three evaluation metrics values are 0.86, 0.85, and 0.77, respectively. The performance of the ALNSynergy model is overall improved by 8.1\%, 9.4\%, and 11.7\% compared with RF. The deep learning model with the best performance is DeepDDS, and its three evaluation metrics values are 0.92, 0.92, and 0.84, respectively. The ALNSynergy model's performance is improved overall by 1.1\%, 1.1\%, and 2.4\% compared to the DeepDDS model.

In addition, the performance of machine learning models in drug synergy prediction problems is worse than that of deep learning models. The deep learning model with the worst performance is DeepSynergy, and its experimental results are better than the best-performing machine learning model RF, which verifies that the deep learning model has powerful fitting ability.

The prediction performance of the DeepSynergy and DeepDDS models improved in turn. DeepSynergy is the first model to introduce deep learning to the task of drug synergy prediction. Its AUC, AUPR, and ACC evaluation metrics values are 0.88, 0.87, and 0.80, respectively. As the most advanced model currently, DeepDDS uses the two-dimensional structure of the drug to learn drug representation and has better prediction performance. Compared with the DeepSynergy model, the improvement values of the DeepDDS model in the AUC, AUPR, and ACC evaluation metrics are 4.5\%, 5.7\%, and 5.0\%, respectively.

However, the above deep learning methods do not consider the alignment between multiple representations. The ALNSynergy model designs a multi-representation alignment function suitable for the field of drug synergy and uses joint optimization to make the alignment function affect the parameters of the ALNSynergy model, so the prediction performance of the ALNSynergy model is better than the above machine learning and deep learning models.

Next, this section further verifies the generalization ability of the ALNSynergy model on the independent dataset provided by AstraZeneca. The experimental results of the ALNSynergy model and the above comparison models on independent dataset are shown in Table 6. As shown in the Table, the predictive performance of the ALNSynergy model is also better than the above-mentioned deep learning models and machine learning models. Specifically, the AUC, AUPR, and ACC values of the ALNSynergy model are 0.66, 0.83, and 0.63, respectively. The prediction performance of the DeepDDS model is also second, with each metric value being 0.65, 0.82, and 0.62, respectively. Compared with the DeepDDS model, the improvement values of the ALNSynergy model under the three metrics are 1.5\%, 1.2\%, and 1.6\%, respectively. The performance comparison results of the remaining models are similar to the benchmark dataset.

According to the above experimental results, the alignment function proposed in this work can improve the representation quality and help improve model performance. At the same time, experimental comparison results with the current most advanced models also verify the excellence of the ALNSynergy model. In addition, experimental results on independent dataset more accurately demonstrate the generalization performance of the ALNSynergy model. This also shows that the ALNSynergy model has a better predictive ability for new drug combinations.

\section{Conclusion}

This work proposes a graph convolutional network with multi-representation alignment, the ALNSynergy model, for predicting drug synergy. The ALNSynergy model proposes a multi-representation alignment function suitable for drug synergy prediction problems, which can reflect the positional relationship between drug combination representations and cell line representations that have a synergistic relationship (antagonism relationship) in the embedding space. In addition, based on the above-mentioned multi-representation alignment function, we further consider the module length of the representation vector. The variation range of the module length is controlled by introducing a temperature coefficient to ensure that the module lengths are of the same magnitude, thereby increasing the accuracy of similar values and accelerating the convergence of the model. Finally, experimental results on the benchmark and independent dataset verify the prediction ability and generalization performance of the ALNSynergy model. We verify the effectiveness of the two innovative elements above through relevant ablation experiments. The experimental comparison with advanced models also demonstrates the highly competitive prediction performance of the ALNSynergy model.

\bibliographystyle{IEEEtran}
\bibliography{reference.bib}
\vspace{-0.8cm}

\end{document}